\documentclass[superscriptaddress,secnumarabic,nobibnotes,aps,prd,showkeys,noshowpacs,onecolumn,10pt]{revtex4-2}%

\usepackage{graphics}
\usepackage{graphicx}
\usepackage{subcaption}
\usepackage{epsf}
\usepackage{bm}
\usepackage{amsmath,amssymb,amsfonts,mathrsfs,amsthm}
\usepackage{latexsym}
\usepackage{enumerate}
\usepackage{comment}
\usepackage[dvipsnames,svgnames,x11names,table]{xcolor}
\usepackage[colorlinks = true,
            linkcolor = Cerulean,
            urlcolor  = magenta,
            citecolor = magenta,
            anchorcolor = NavyBlue]{hyperref}
\usepackage{epstopdf}%
\usepackage{bbm}
\def\be{\begin{equation}}
\def\ee{\end{equation}}

\makeatother

\begin{document}

\title{Compact Stars in Symmetric Teleparallel Scalar-Tensor Gravity}
\author{Grigorios Panotopoulos}
\email{grigorios.panotopoulos@ufrontera.cl}
\affiliation{Departamento de Ciencias F\'{\i}sicas, Universidad de la Frontera, Casilla
54-D, 4811186 Temuco, Chile}
\author{Andr\'es Lueiza-Colip\'i}
\email{a.lueiza01@ufromail.cl}
\affiliation{Departamento de Ciencias F\'{\i}sicas, Universidad de la Frontera, Casilla
54-D, 4811186 Temuco, Chile}
\author{Nikolaos Dimakis}
\email{nikolaos.dimakis@ufrontera.cl}
\affiliation{Departamento de Ciencias F\'{\i}sicas, Universidad de la Frontera, Casilla
54-D, 4811186 Temuco, Chile}
\author{Andronikos Paliathanasis}
\email{anpaliat@phys.uoa.gr}
\affiliation{Institute of Systems Science, Durban University of
Technology, Durban 4000, South Africa}
\affiliation{Centre for Space Research, North-West University, Potchefstroom 2520, South Africa}
\affiliation{National Institute for Theoretical and Computational Sciences (NITheCS), South Africa}
\affiliation{Departamento de Matem\'{a}ticas, Universidad Cat\`{o}lica del Norte, Avda.
Angamos 0610, Casilla 1280 Antofagasta, Chile}

\begin{abstract}\noindent
We investigate the existence of static, spherically symmetric compact objects within the framework of symmetric teleparallel scalar–tensor gravity. This theory extends the Brans-Dicke and scalar-tensor models within the symmetric teleparallel formalism. We consider a nontrivial connection that allows for genuinely nontrivial solutions in the limit of General Relativity. The field equations admit a minisuperspace description and by applying the method of variational symmetries we construct the corresponding conservation laws in vacuum. The application of these conservation laws enables the reconstruction of analytic black-hole solutions. Finally, we study the interior structure of compact objects matched to an extremal Reissner-Nordstr\"{o}m exterior and show that the symmetric teleparallel scalar-tensor theory supports the existence of viable astrophysical objects.
\end{abstract}

\maketitle

%%%%%%%%%%%%%%%%%%%%%%%
\section{Introduction}
%%%%%%%%%%%%%%%%%%%%%%%

General Relativity (GR) is a well-tested gravitational theory for describing
local gravitational phenomena. However, it fails to account by itself for the recent cosmological observations, which indicate that the universe is currently
undergoing an accelerated expansion. Furthermore, it does not provide a mechanism to resolve the existing cosmological tensions \cite%
{Teg,Kowal,Komatsu,suzuki11,ade18,cco1,yy1,desi4,dd,cosm}. Various modifications to the Einstein-Hilbert action have been constructed, many of which introduce effective corrections that manifest due to possible quantum effects \cite%
{qc1,qc2,qc3,qc4}. As a result, new directions in gravitational research have been opened, leading to the development of what is referred as extended or modified theories of gravity. These theories introduce additional geometrodynamical degrees of freedom meant to account for the observable phenomena \cite{md1,md2b,md2,md3,md4}. 

The geometric framework of GR is defined by the metric
tensor and the associated Levi-Civita connection, which is used to construct the Riemannian curvature. However, this is not the only geometric structure capable of reproducing the result obtained by Einstein's field equations. Within the
framework of teleparallel gravity, the torsion scalar, defined by the vierbein field, serves as the analogue of the Ricci scalar appearing in the gravitational action, leading to the Teleparallel Equivalent of General Relativity (TEGR) \cite{ein28,Hayashi79}. In conjunction to this, a dynamically equivalent theory can be devised in terms of the nonmetricity scalar, as obtained from a symmetric and flat connection. This approach leads to the Symmetric Teleparallel Equivalent of General Relativity (STEGR) \cite{Nester:1998mp}. The equivalence among these gravitational theories, based on linear expressions of the relevant scalars, is broken when nonlinear correction terms are introduced in the gravitational
action \cite{Ferraro,Koivisto2,Koivisto3}, or when scalar fields are nonminimally coupled to gravity \cite{te2,te1,sc1,sc2}.

For a gravitational theory to be physically viable, it must not only account for the large-scale structure of the universe but also describe local gravitational phenomena and thus recover GR in the appropriate limit. In this study, we investigate the existence of static, spherically symmetric solutions that can describe compact astrophysical objects within the framework of symmetric teleparallel scalar-tensor gravity \cite{sc1,sc2,sc3}%
. This theory generalizes the Brans-Dicke and scalar-tensor models \cite {bd1,bd2} within the symmetric teleparallel formalism, where the presence of a scalar field is essential for defining the physical theory. Moreover, the theory is consistent with Mach's principle. In this gravitational model, the scalar field is nonminimally coupled to the nonmetricity scalar. Let us mention that, symmetric
teleparallel $f\left( Q\right) $-gravity is included in this general
framework as a special case \cite{Koivisto2,Koivisto3}. 

There exists an extensive literature investigating the symmetric
teleparallel scalar-tensor and $f\left( Q\right) $ theories as possible dark energy
candidates used in the description of cosmic acceleration \cite%
{fq1,fq2,fq3,fq4,fq5,fq6,fq7,fq8,fq9,fq10}. However, there are limited
studies on the description of local gravitational phenomena. Vacuum static spherically symmetric solutions in $f\left( Q\right) $-gravity have been investigated in \cite{bh1,bh2a,bh2b,bh2}, while within the framework of symmetric teleparallel scalar-tensor theory analytic solutions have been
derived in \cite{bh2,Baha1}. Recently, in \cite{starfQ}, relativistic stars within the $f\left( Q\right) $-gravity framework, which admit the de Sitter-Schwarzschild solution as an attractor, were investigated. Although numerous studies exist in the
literature on similar topics, most of them focus on
the case of STEGR, or consider a symmetric and teleparallel connection that does not recover the GR limit, providing a non-physically
viable theory; for an extended discussion see \cite{starfQ} and \cite{telg}. 

In the following, we extend the analysis presented in \cite{starfQ} by
studying the existence of compact objects in symmetric teleparallel
scalar-tensor gravity. We consider a connection defined in the noncoincident
gauge, which provides a set of field equations that allow for nontrivial
solutions. We apply the method of variational symmetries \cite{sym1,sym2}, to determine
conservation laws, on the base of which, we chose the particular theory under consideration. The presence of the relevant symmetries is connected to the existence of known closed-form
vacuum solutions. We obtain interior solutions of relativistic stars made of isotropic matter. The solutions possess as an exterior vacuum solution a geometry that resembles the extremal Reissner-Nordstr\"{o}m (RN) geometry, although in this work we consider compact objects that are electrically neutral. The structural properties as well as their stability based on the Harrison-Zel'dovich Novikov criterion are discussed. Moreover, we compare to a number of current astrophysical constraints, which are the following: i) The gravitational wave signal from the {\bf GW}190814 event, observed by LIGO \cite{LIGOScientific:2020zkf}. The binary system consists of a black hole and a companion star, the mass of which is found to be $2.50 M_{\odot}$ - $2.67 M_{\odot}$, higher than stellar masses of conventional neutron stars. ii) The supernova remnant HESS J1731-347 compact object that is characterized by a strangely lower mass $M = 0.77^{+0.20}_{-0.17} M_{\odot}$ and radius $R = 10.4^{+0.86}_{-0.78} km$ than usual \cite{Doroshenko:2022nwp, Rather:2023tly}. iii) NICER results, such as PSR J0030+0451 \cite{Riley:2019yda, Miller:2019cac}, PSR J0437-4715 \cite{Bogdanov:2012md, Choudhury:2024xbk} and PSR J0740+6620 \cite{Riley:2021pdl, Miller:2021qha}. iv) The most massive pulsars at around two solar masses, PSR J0348+0432 \cite{Antoniadis:2013pzd, Zhao:2015tra} and PSR J1614-2230 \cite{Demorest:2010bx, Lenzi:2012xz}.

The structure of the paper is as follows: In Section II, we introduce the gravitational model under consideration,
namely the scalar--tensor theory formulated within the symmetric
teleparallel framework. The geometric structure of static, spherically
symmetric spacetimes in symmetric teleparallel gravity is discussed in
Section III, where we also provide a detailed analysis for our choice of connection, to ensure that the resulting gravitational theory is not constrained to a GR equivalent description. Moreover, we examine the variational symmetry structure of the problem and distinguish the class of theories that allow for their emergence. Next, vacuum solutions in the absence of matter are presented in the fourth section. Section V contains the main analysis of this work, where we investigate the
existence of compact stars within the symmetric teleparallel theory of
gravity. We show that there exists a viable class of interior solutions
sourced by a physically acceptable fluid, which continuously match the exterior extremal Reissner-Nordstr\"{o}m solution at the stellar surface. The compactness, gravitational redshift, and mass of the star are computed numerically. Finally, in Section VI, we summarize our work and draw our conclusions.

%%%%%%%%%%%%%%%%%%%%%%%%%%%%%%%%%%%%%%%
\section{Action of the theory and field equations}
%%%%%%%%%%%%%%%%%%%%%%%%%%%%%%%%%%%%%%%

We consider the symmetric teleparallael scalar-tensor theory of gravity with action integral \cite{sc1,sc2,sc3}
\begin{equation}\label{action}
   S =\frac{1}{\kappa}\int\!\! d^4x \sqrt{-g} \left[ \frac{ A(\phi)}{2} Q -\frac{B(\phi)}{2} \partial_\mu \phi \partial^\mu \phi - V (\phi) \right] + S_m,
\end{equation}
where $\kappa=8\pi G$ (we work in $c=1$ units) and $Q$ is the nonmetricity scalar. 

Gravitation is purely governed by the nonmetricity in the sense that we consider a flat and symmetric connection $\Gamma^{\alpha}_{\;\mu\nu}$, that is independent from the metric. The curvature and the torsion are by construction zero  
\begin{align}
  & R^{\lambda}_{~\mu\nu\kappa} = \frac{\partial}{\partial x^\nu} \Gamma_{~\mu\kappa }^{\lambda } -\frac{\partial}{\partial x^\kappa} \Gamma_{~\mu \nu}^{\lambda } + \Gamma_{~\mu \kappa }^{\eta }\Gamma_{~\nu\eta}^{\lambda } -
    \Gamma_{~\mu \nu }^{\eta } \Gamma_{~\kappa\eta}^{\lambda } =0, \\
  &\mathcal{T}_{~\mu \nu }^{\alpha }=2\left( \Gamma _{~\mu \nu }^{\alpha }-\Gamma _{~\nu\mu }^{\alpha }\right) =0.
\end{align}
As seen from Eq. \eqref{action}, the theory is in principle nonminimally coupled to a scalar field $\phi$. The STEGR, with or without a scalar field, is recovered when the function $A(\phi)$ becomes a constant. At this point let us emphasize that this theory is mapped to $f(Q)$ gravity in the case where there is no kinetic term for the scalar field, i.e. when $B(\phi)=0$.

The theory is characterized by three fundamental fields: the metric $g_{\mu\nu}$, the connection $\Gamma^{\alpha}_{\;\mu\nu}$ and the scalar field $\phi(r)$. Variation of the action \eqref{action} with respect to the metric produces the field equations \cite{Baha1}
\begin{equation} \label{fieldmet}  
   A(\phi) G_{\mu\nu} + 2 A_\phi(\phi) P^\alpha_{\;\mu\nu} \partial_\alpha \phi + \frac{1}{2} g_{\mu\nu} \left( B(\phi) \partial_\mu \phi \partial^\mu \phi +2 V(\phi) \right) - B(\phi) \partial_\mu \phi \partial_\nu \phi = \kappa \, T_{\mu\nu}.
\end{equation}
In the above expression, the $G_{\mu\nu}$ is Einstein's tensor and the $T_{\mu\nu}$ stands for the energy momentum tensor of the matter content $S_m$. The index $\phi$ is used to denote derivation with respect to the scalar field, i.e. $A_\phi =\frac{dA}{d\phi}$. 

The tensor $P^\alpha_{\;\mu\nu}$ is referred as the superpotential and it is defined as
\begin{equation}
P^\alpha_{\phantom{\alpha}\mu\nu} = -\frac{1}{4} Q^{\alpha}_{\phantom{\alpha}%
\mu\nu} + \frac{1}{2} Q_{(\mu\nu)}^{\phantom{(\mu\nu)}\alpha} + \frac{1}{4}
\left( Q^\alpha - \widetilde{Q}^\alpha \right)g_{\mu\nu} - \frac{1}{4}
\delta^{\alpha}_{\phantom{\alpha}(\mu}Q_{\nu)},
\end{equation}
where $Q_{\alpha\mu\nu}=\nabla_\alpha g_{\mu\nu}$ is the nonmetricity tensor, with  $Q_\alpha=Q_{\alpha\phantom{\mu}\mu}^{\phantom{\alpha}\mu}$ and $\widetilde{Q}_\alpha = Q^{\mu}_{\phantom{\mu}\alpha\mu}$. 

The definition of the nonmetricity scalar arises from the contraction between the nonmetricity tensor and the corresponding superpotential, that is,
\begin{equation} \label{Qscdef}
  Q = Q_{\alpha\mu\nu}P^{\alpha\mu\nu} .
\end{equation}
Let us note that with the above definition of the nonmetricity scalar, as seen in Eq. \eqref{Qscdef}, the scalar field $\phi$ is canonical if $A(\phi)$ and $B(\phi)$ have the same sign and phantom in the opposite case. 

Advancing to the second independent field of the theory, the variation of the action with respect to the connection, $\Gamma^{\alpha}_{\;\mu\nu}$, yields \cite{Baha1}
\begin{equation} \label{fieldgamma}
\nabla_\mu \nabla_\nu \left(\sqrt{-g} A(\phi) P^{\mu\nu}_{\phantom{\mu\nu}\alpha} \right) =0 .  
\end{equation} 
This equation describes the law of motion for the geometrodynamical degrees of freedom introduced by the connection. 

Finally, the modified Klein-Gordon type of equation, that is the field equation for $\phi$, reads
\begin{equation} \label{fieldphi}
   2 B(\phi) \tilde{\nabla}_\mu \tilde{\nabla}^\mu \phi + B_\phi(\phi) \partial_\mu \phi \partial^\mu \phi + A_\phi(\phi) Q - 2 V_\phi(\phi) = 0.
\end{equation}
We use $\tilde{\nabla}$ here to denote the covariant derivative with respect to the Levi-Civita connection. Eqs. \eqref{fieldmet}, \eqref{fieldgamma}, and \eqref{fieldphi} constitute the complete system of gravitational field equations, which must be satisfied in order to obtain solutions within the scalar–nonmetricity theory. We want to underline that all equations of this set are equally important. Because, in many cases, the field equation for the connection is ignored in the literature. In order for a consistent solution to be derived, there must be used a  connection which, in conjunction with the metric and the scalar field, satisfies Eq. \eqref{fieldgamma}; for a relative discussion see \cite{starfQ}.

%%%%%%%%%%%%%%%%%%%%%%%%%%%%%%%%%%%%%%%%%%%%%%%%%%%%%%%%%%%%%%%%%%%%%%
\section{Static, and spherically symmetric spacetimes: Structure equations}
%%%%%%%%%%%%%%%%%%%%%%%%%%%%%%%%%%%%%%%%%%%%%%%%%%%%%%%%%%%%%%%%%%%%%

At this point, we introduce a static and spherically symmetric spacetime, whose line element is given by
\begin{equation}\label{genmetric}
  ds^2 = - a(r)^2 dt^2 + n(r)^2 dr^2 + b(r)^2 \left( d\theta^2 + \sin^2 \theta d\varphi^2 \right).
\end{equation}
We additionally assume that the scalar field is static, having a pure $r$-dependence, i.e. $\phi=\phi(r)$. The requirement that the connection is flat, symmetric, and that it inherits the $SO(3) \times \mathbb{R}$ isometries of the line element \eqref{genmetric}, leads to two distinct families of connections \cite{bh1,Hohmann}. The two families have been used in various contexts in the derivation of static, spherically symmetric solutions in theories with nonmetricity \cite{bh1,bh2,Baha1,starfQ,Hohmann}. 

In this work, we consider the connection belonging to the second family, which has the nonzero components
\begin{equation} \label{consol1}
\begin{split}
& \Gamma _{\;tt}^{t}=c_{1}+c_{2}-c_{1}c_{2}\gamma _{1},\quad \Gamma
_{\;tr}^{t}=\frac{c_{2}\gamma _{1}(c_{1}\gamma _{1}-1)}{\gamma _{2}},\quad
\Gamma _{\;rr}^{t}=\frac{\gamma _{1}(1-c_{1}\gamma _{1})(c_{2}\gamma
_{1}+1)-\gamma _{2}\gamma _{1}^{\prime }}{\gamma _{2}^{2}}, \\
& \Gamma _{\;\theta \theta }^{t}=-\gamma _{1},\quad \Gamma _{\phi \phi
}^{t}=-\sin ^{2}\theta \gamma _{1},\quad \Gamma
_{\;tt}^{r}=-c_{1}c_{2}\gamma _{2},\quad \Gamma _{\;tr}^{r}=c_{1}\left(
c_{2}\gamma _{1}+1\right) , \\
& \Gamma _{\;rr}^{r}=\frac{1-c_{1}\gamma _{1}(c_{2}\gamma _{1}+2)-\gamma
_{2}^{\prime }}{\gamma _{2}},\quad \Gamma _{\;\theta \theta }^{r}=-\gamma
_{2},\quad \Gamma _{\;\phi \phi }^{r}=-\gamma _{2}\sin ^{2}\theta ,\quad
\Gamma _{\;t\theta }^{\theta }=c_{1}, \\
& \Gamma _{\;r\theta }^{\theta }=\frac{1-c_{1}\gamma _{1}}{\gamma _{2}}%
,\quad \Gamma _{\phi \phi }^{\theta }=-\sin \theta \cos \theta ,\quad \Gamma
_{t\phi }^{\phi }=c_{1},\quad \Gamma _{\;r\phi }^{\phi }=\frac{1-c_{1}\gamma
_{1}}{\gamma _{2}},\quad \Gamma _{\;\theta \phi }^{\phi }=\cot \theta .
\end{split}
\end{equation}%
The $c_{1}$, $c_{2}$ are constants, while the $\gamma _{1}(r)$, $\gamma _{2}(r)$ are functions of $r$. We use the prime to denote differentiation with respect to the latter. The aforementioned connection coefficients are meant to be constrained by the equations of motion, and especially that of the connection, Eq. \eqref{fieldgamma}. 

The reason behind the selection of the connection belonging to the second family, with components given by Eqs. \eqref{consol1}, is the following: Both of the previously mentioned families of connections lead to a non-zero off-diagonal component in the metric field equations \eqref{fieldmet}. In the case of the first family, the only way to satisfy the field equations - assuming a diagonal energy momentum tensor - is by reducing the theory to STEGR in order to trivialize the non-diagonal component. In the case of the connection we consider here, the off-diagonal component reads
\begin{equation}
   (2 c_1 c_2 \gamma_1-2 c_1-c_2)\phi '(r) A_\phi(\phi)=0,
\end{equation}
which allows for more possibilities. The choice of either $A_\phi=0$ or $\phi '=0$ inevitably reduces the theory to that of STEGR with or without a cosmological constant, which is dynamically equivalent to GR. However, this can be avoided by setting the combination in the parenthesis equal to zero. With this choice, we can study a genuinely modified theory of gravity. The simplest scenario, and the one which we follow here, is to invoke $c_1=0=c_2$, for more details we refer the reader to \cite{bh2}.

With the previous considerations, and for a general anisotropic fluid energy-momentum tensor, $T_{\mu\nu}=\mathrm{diag}(-\rho,p_r,p_t,p_t)$, the field equations for the metric \eqref{fieldmet}, reduce to
\begin{subequations} \label{thefluid}
\begin{align} \nonumber
 \kappa \rho =  &  \frac{\phi'}{2 b^2 n^2 \gamma_2} \left(4 b \gamma_2 b' A_\phi(\phi)+b^2 \left(\gamma_2 B(\phi)\phi'-2 A_\phi(\phi)\right)-2 n^2 \gamma_2^2 A_\phi(\phi)\right) \\
  & +\frac{A(\phi)}{b^2 n^3} \left(-2 b b' n' + n \left(2 b b''+b'^2\right)-n^3\right)+V(\phi)  \\
  \kappa p_r = & \frac{A(\phi) \left(2 b a' b'+  a \left(b'^2-n^2\right)\right)}{a b^2 n^2}+\frac{1}{2} \phi ' \left(\frac{2 A_\phi(\phi)}{n^2\gamma_2}-\frac{B(\phi)\phi '}{n^2}-\frac{2 \gamma_2 A_\phi(\phi)}{b^2}\right)+V(\phi) \\ \nonumber
  \kappa p_t =  & \frac{1}{2 a b n^3 \gamma_2} \Big[2 \gamma_2 \left(b n a' \phi ' A_\phi(\phi)+A(\phi) \left(b n a''+n a' b'-b a' n'\right)\right) \\
  & +2 a \gamma_2 \left(n b' \phi ' A_\phi(\phi)+A(\phi) \left(n b''-b' n'\right)\right) \nonumber \\
  & + a b n \left(\phi ' \left(\gamma_2 B(\phi)\phi '-2 A_\phi(\phi)\right)+2 n^2 \gamma_2 V(\phi)\right) \Big]  .
\end{align}
\end{subequations}
The connection equation, Eq. \eqref{fieldgamma}, becomes
\begin{equation} \label{fieldgamma2}
  \begin{split}
    & 2 n \left(b^2-n^2 \gamma_2^2\right) \left(a' \phi ' A_\phi(\phi)+a \phi '^2 A_{\phi\phi}(\phi)+a(r) \phi '' A_{\phi}(\phi)\right) \\
    & +4 a n \phi ' A_{\phi}(\phi) \left(b b'-n^2 \gamma_2 \gamma_2'\right)-2 a n' \phi ' A_{\phi}(\phi) \left(b^2+n^2 \gamma_2^2\right)=0,
  \end{split}
\end{equation}
while that for the relevant equation for the scalar field, Eq. \eqref{fieldphi}, leads to
\begin{equation} \label{fieldphi2}
  \begin{split}
     & a b^2 n\left( 2  B(\phi) \phi'' + B_\phi(\phi) \phi '^2 - 2 n^2 V'(\phi) \right) + 2 b B(\phi) \left(b \left(n a'-a  n'\right)+2 a n b'\right)\phi' - \\
     & \frac{2 A_\phi(\phi)}{\gamma_2^2} \Big[ n \gamma_2 a' \left(n^2 \gamma_2^2 -2 b \gamma_2 b'+b^2\right)-n a \left(b^2 \gamma_2'-2 \gamma_2  b b'+\gamma_2^2 b'^2\right) \\
     & - a b^2 \gamma_2 n'+ an^2 \gamma_2^3 n'+a n^3 \gamma_2^2 \left(\gamma_2'-1\right) \Big] =0.
   \end{split}
\end{equation}
Note that the equation of the connection, Eq. \eqref{fieldgamma}, has a particular solution given by \cite{Baha1}
\begin{equation} \label{trivcon}
  \gamma_2 = \frac{b}{n}.
\end{equation}
A straightforward substitution of this $\gamma_2$ into Eq. \eqref{fieldgamma} immediately satisfies  the latter. This exact choice for the function $\gamma_2$ also works in the case of static, spherically configurations in $f(Q)$ theory \cite{starfQ}.

%%%%%%%%%%%%%%%%%%%%%%%%%%
\section{Solutions in the absence of matter content}
%%%%%%%%%%%%%%%%%%%%%%%%%%

In order to motivate the choice of a particular theory inside this infinite class of scalar nonmetricity theories, let us study the symmetry structure of the relevant vacuum system. To this end, note that the vacuum field equations ($\rho=p_r=p_t=0$) can be generated from the minisuperspace Lagrangian
\begin{equation}\label{miniLag}
  \begin{split}
   L = & \frac{1}{n}\left( 2 b A(\phi )a'  b'+a A(\phi) b'^2-\frac{a b^2 B(\phi) \phi '^2}{2} + a b^2  A_\phi(\phi ) \psi ' \phi '\right) \\
   & + n \left( \frac{a  A_\phi(\phi ) \phi'}{\psi '} + a A(\phi)-a b^2 n V(\phi) \right),
   \end{split}
\end{equation}
where we introduced an auxiliar scalar field $\psi$ defined by the relation $\gamma_2=1/\psi'$. Note, that due to the transformation law of the connection, the $\psi$ is a geometric scalar and it expresses the extra degree of freedom owed to the connection. There is an active debate in what regards the number of the extra degrees of freedom encountered in nonmetricity theories \cite{df1,df2,df3}. 

Constructing an equivalent Lagrangian formulation of a system with finite degrees of freedom simplifies the use of variational symmetries in order to obtain conservation laws. As is well known, the latter are essential in order to simply the gravitational field equations or to construct invariant solutions. 

Without entering in details about the Noether symmetry formalism and how it is applied in the case of gravitational Lagrangians - a subject well established in the literature \cite{sym1,sym2} - let us note that the action with Lagrangian function \eqref{miniLag} has the obvious symmetry $\frac{\partial}{\partial \psi}$. The reason for this is because Eq. \eqref{miniLag} is trivially invariant under translations in $\psi$ since there is no explicit dependence on the latter, but only on its first derivative. 

Additional, and not so trivial symmetries, can be obtained for specific expressions of the involved scalar field functions $A$, $B$ and $V$. Two supplementary symmetries emerge, with generators
\begin{align}
  X_1 & = a \frac{\partial}{\partial a} - \frac{A(\phi)}{A_\phi(\phi)} \frac{\partial}{\partial \phi}, \\
  X_2 & = n \frac{\partial}{\partial n} - \frac{a}{2} \frac{\partial}{\partial a}+ b \frac{\partial}{\partial b} - \frac{A(\phi)}{2A_\phi(\phi)} \frac{\partial}{\partial \phi}
\end{align}
under the conditions
\begin{equation}
\label{the1}
  A= A(\phi), \quad B= B_0 \frac{A_\phi(\phi)^2}{A(\phi)}, \quad V= 0.
\end{equation}
The resulting conserved charges are
\begin{align}
  I_1 & = \frac{a \left(B_0 b^2 \gamma_2 \phi' A_\phi-A(\phi) \left(b^2-2 b\gamma_2 b'+n^2 \gamma_2^2\right)\right)}{n \gamma_2} ,\\
  I_2 & = \frac{4 b^2 \gamma_2 a' A(\phi)+a \left(B_0 b^2 \gamma_2 \phi' A_\phi-A(\phi) \left(b^2-2 b \gamma_2 b'+n^2 \gamma_2^2\right)\right)}{2 n \gamma_2},
\end{align}
where we have performed the substitution $\psi'=1/\gamma_2$.

In the case of a non-vanishing potential, $V(\phi)\neq 0$, we obtain a single additional symmetry in the form of
\begin{equation}
  X_3 = \left(\mu-5\right) X_1 + 2 \left(\mu-1 \right)X_2
\end{equation}
subjected to the conditions for the theory
\begin{equation}
  A= A(\phi), \quad B= B_0 \frac{A_\phi(\phi)^2}{A(\phi)}, \quad V= V_0 A(\phi)^\mu,
\end{equation}
where $A(\phi)$ is an arbitrary function of $\phi$ and $B_0$, $V_0$ and $\mu$ are constants. The conserved charge being of course in this case
\begin{equation}
  I_3 = \left(\mu-5\right) I_1 + 2 \left(\mu-1 \right)I_2 .
\end{equation}

We observe that in both cases, $B(\phi)$ is subject to a specific relation with respect to $A(\phi)$. At a first glance it may seem that there exists an infinite set of theories admitting symmetries, since $A(\phi)$ remains undetermined. However, this is not the case, due to the freedom of reparameterizing the scalar field. To demonstrate this, we bring the scalar field into canonical form, that is, we perform a transformation $\phi\rightarrow \chi$ which renders the function in front of the scalar kinetic term of the action \eqref{action} a constant. We thus demand
\begin{equation}
  B(\phi)^{1/2} d \phi = B_0^{1/2} \frac{A_\phi(\phi)}{A(\phi)^{1/2}}  d \phi =  B_0^{1/2}  d\chi . 
\end{equation}
The mapping $\phi= A^{-1}(\chi/4)$, leads to the expressions
\begin{equation}
  A \sim \chi^2, \quad B= B_0, \quad V \sim \chi^{2\mu}.
\end{equation}
It is thus only a quadratic non-minimal coupling of the scalar field with the nonmetricity scalar $Q$ that gives rise to extra conserved quantities, with or without a (power-law) potential.

The existence of this first order equation explains to a large extent the solutions obtained previously in \cite{Baha1} for the vacuum case. The simplest of them being
\begin{equation} \label{RNmet}
  ds_1^2 = -\left(1-\frac{M}{r}\right)^2 dt^2 + \left(1-\frac{M}{r}\right)^{-2} dr^2 + r^2 \left( d\theta^2 + \sin^2 \theta d\varphi^2 \right),
\end{equation}
with $M$ being a constant of integration that may be interpreted as the mass of the object that generates the gravitational field. The geometry is an extremal Reissner–Nordstr\"om black hole with the mass equal to the ``charge'' since $(1-M/r)^2=1-2M/r+M^2/r^2$. This solution is obtained for the theory with
\begin{equation}
\label{a0}
  A(\phi)=\frac{A_0}{8} \phi^2, \quad B(\phi)=-A_0, \quad V(\phi)=0, \quad \phi = \phi_0 \left(1-\frac{M}{r}\right)^{-1/2}, \quad \gamma_2 = r-M .
\end{equation}
Note that the function of the connection $\gamma_2$ is of the type \eqref{trivcon}, which automatically trivializes Eq. \eqref{fieldgamma2}. The metric \eqref{RNmet} is the solution that we shall adopt for the exterior of the stellar configuration we present in this work. From Eq. \eqref{a0} we infer that the latter solution does not exist in the limit of $f(Q)$-gravity since the potential is zero.  

Another interesting solution, also previously derived in \cite{Baha1}, is given in terms of
\begin{equation}
  ds_2^2 = -\left[\left(\frac{8 A_0+B_0}{8 A_0-B_0}\right)^2 -m r^{\frac{8 A_0-B_0}{8 A_0+B_0}} \right] dt^2 + \left[\left(\frac{8 A_0+B_0}{8 A_0-B_0}\right)^2 -m r^{\frac{8 A_0-B_0}{8 A_0+B_0}} \right]^{-1} dr^2 + r^2 d\Omega^2,
\end{equation}
with
\begin{equation}
  A(\phi)=A_0 \phi^2, \quad B(\phi)=B_0, \quad V(\phi)=0, \quad \phi = \phi_0 r^{-\frac{8 A_0}{8 A_0+B_0}}, \quad \gamma_2= \frac{8 A_0+B_0}{8 A_0-B_0}r .
\end{equation}
The function $\gamma_2$ here is not of the type given by Eq. \eqref{trivcon} and it forms a distinct solution. In the limit where $B_0=0$, the spacetime is reduced to the Schwarzschild metric.

%%%%%%%%%%%%%%%%%%%%%%%
\section{Interior solutions of Quark stars: The case of the simplified MIT bag model}
%%%%%%%%%%%%%%%%%%%%%%%

In this section, we focus on the investigation of compact objects within the scalar–nonmetricity theory, considering the vacuum solutions discussed earlier as limiting cases, such that the resulting configurations describe relativistic stars. We adopt the class of theories \eqref{the1} with a vanishing potential, where without loss of generality we may assume
\begin{equation}
\label{the2}
   A = A_0 \phi^2, \quad B= B_0, \quad V(\phi) = 0 ,
\end{equation}
where both constant parameters $A_0, B_0$ are dimensionless.
For a general nonconstant function $B(\phi)$, one can always introduce a new scalar field through an appropriate change of variables such that the model \eqref{the1} is written in the form of \eqref{the2}.

Now that we have left the symmetry analysis behind, we proceed to fix the gauge by setting the $b(r)$ of the metric \eqref{genmetric} equal to the radial distance, i.e. $b=r$. At the same time, we parameterize the other two functions in terms of two metric potentials $\nu(r)$ and $\lambda(r)$ as
\begin{equation}
  a = e^{\nu(r)/2}, \quad n= e^{\lambda(r)/2} .
\end{equation}
For the function $\gamma_2$, which enters the connection, we adopt the solution \eqref{trivcon}. As previously stated, the latter satisfies the Eq. \eqref{fieldgamma2} of the connection. In this parametrization, the function $\gamma_2$ reads
\begin{equation} \label{connectionsol}
   \gamma_2 = r \, e^{-\lambda (r)/2},
\end{equation}
and the non-metricity scalar assumes the expression
\begin{equation}\label{nonmetsol}
  Q = \frac{2 e^{-\lambda}}{r^2} \left(e^{\frac{\lambda}{2}}-1\right) \left(e^{\frac{\lambda}{2}}-r \nu '-1\right) .
\end{equation}

For the fluid matter source, the energy density and pressure components are subsequently obtained from Eqs. \eqref{thefluid}
\begin{subequations} \label{fluid}
\begin{align}
  \rho & = \frac{e^{-\lambda}}{2 \kappa  r^2} \left[ 2 A_0 \phi^2 \left(r \lambda '+e^{\lambda}-1\right)+8 A_0 r \left(e^{\lambda/2}-1\right) \phi \phi '-B_0 r^2 \phi'^2 \right] , \\
  p_r & =  \frac{e^{-\lambda}}{2 \kappa  r^2} \left[ 2 A_0 \phi^2 \left(1-e^{\lambda }+r \nu '\right)- B_0 r^2 \phi '^2 \right] , \\
  p_t & = \frac{e^{-\lambda}}{4\kappa  r} \left[ A_0 \phi ^2 \left(2 r \nu ''+r \nu'^2+2 \nu'-\lambda ' \left(r \nu '+2\right)\right)-4 A_0 \phi \phi' \left(2 e^{\frac{\lambda}{2}}-r \nu'-2\right)+2 B_0 r \phi'^2\right] .
\end{align}
\end{subequations}
With the above expressions, it is easily verified that the metric field equations \eqref{fieldmet} are satisfied. The connection equation is already solved by \eqref{connectionsol}. We are thus left with the scalar field Eq. \eqref{fieldphi2}, which now reads
\begin{equation}
   4 A_0 \left(e^{\lambda/2}-1\right) \left(e^{\lambda/2}-r \nu '-1\right)\phi + B_0 r \left(\phi ' \left(r \nu '-r \lambda '+4\right)+2 r \phi''\right) =0 .
\end{equation}
Integration of the above with respect to $\nu$ leads to
\begin{equation}\label{nusol}
   \nu = \int \! \frac{B_0 r \left(\left(r \lambda '-4\right) \phi '-2 r \phi ''\right)-4 A_0 \left(e^{\lambda/2}-1\right)^2 \phi}{r \left(B_0 r \phi'-4 A_0 \left(e^{\lambda/2}-1\right) \phi\right)} dr.
\end{equation}
Of course, performing this integration introduces an additive constant in $\nu$, which appears multiplicatively in the temporal component of the metric. This constant can be used to impose continuity between the interior and exterior solutions. We remark that in the expressions for the fluid energy density and the pressure components \eqref{fluid}, only the derivative $\nu'$ appears. As a result, it is not necessary to explicitly calculate the integral \eqref{nusol} for the study of the fluid. 

One possible approach at this point would be to choose specific functions $\lambda$, $\phi$ and study the resulting fluid expressions. However, it is not straightforward to determine which choices would correspond to physically reasonable matter content. Therefore, the metod we adopt here is to assume a specific equation of state and use it as a basis to construct the interior solution. In what follows, we outline the basic steps of this procedure.

First, we impose the condition for an isotropic star, in other words, we demand
\begin{equation}
  p_t = p_r = p .
\end{equation}
Subsequently, we introduce the linear equation-of-state
\begin{equation}\label{eqofstate}
p= \frac{1}{3}\left(\rho -4 \mathcal{B}\right),
\end{equation}
representing the simplified quark matter MIT bag model \cite{qstar1,qstar2}, where $\mathcal{B}$ represents a vacuum constant pressure. The continuity equation $\widetilde{\nabla}_\mu T^{\mu\nu}=0$ then implies,
\begin{equation} \label{eqcont}
   p'+\frac{1}{2} (p+\rho ) \nu ' =0 .
\end{equation}
Note that the left hand side of Eq. \eqref{fieldmet} has a zero covariant divergence with respect to $\widetilde{\nabla}_\mu$ if all field equations are satisfied, including that of the connection \eqref{fieldgamma2}. This is the reason why it is imperative to use a $\gamma_2$ function that satisfies the equation of motion for the connection; which is exactly the case of Eq. \eqref{connectionsol}. We further set
\begin{equation}
  B_0 = - 8 A_0 ,
\end{equation}
so that we implement the theory that admits as an external solution that of the extremal Reissner-Nordstr\"om, see Eq. \eqref{RNmet}. 

We then reparameterize the metric potential $\lambda(r)$ with respect to the mass function $m(r)$ as
\begin{equation}
  e^{\lambda(r)}= \left(1-\frac{m(r)}{r} \right)^{-2} ,
\end{equation}
so that at the radius $R$ of the star we have $m(r)=M$ and the $g_{rr}$ component of the internal metric is equal to the corresponding component of the external solution \eqref{RNmet}; thus ensuring continuity. Remember that, for the $g_{tt}$ component, continuity is assured due to the constant of integration included in Eq. \eqref{nusol}. The radius of the star, $R$, is of course defined at the value of the first positive root of the equation formed by the pressure becoming zero, $p(R)=0$. 

With the previous considerations we are left with three differential equations: one for the mass, 
\begin{equation} \label{masseq}
  \begin{split}
    m' =  & \frac{1}{2 A_0 r (r-m) \phi^2} \Bigg[ 4 r^4 \left(8 \pi  \mathcal{B}-A_0 \phi '^2\right)+4 A_0 r^2 m\, \phi ' \left(2 r \phi '-\phi \right)-A_0 m^2 \left(\phi-2 r \phi '\right)^2\\
    & +24 \pi  r^4 p \Bigg],
  \end{split}
\end{equation}
another for the pressure,
\begin{equation}\label{preseq}
  \begin{split}
    p' =&  \frac{1}{A_0 r (r-m)^2 \phi^2} \Bigg[ 2 (B+p) \Big(A_0 \left(m^2 \left(4 r^2 \phi '^2+\phi (r)^2\right)-2 r m \left(4 r^2 \phi '^2+\phi^2\right)+4 r^4 \phi'^2\right)\\
    &-8 \pi  r^4 p\Big) \Bigg],
  \end{split}
\end{equation}
and finally for the scalar field,
\begin{equation}\label{phieq}
  \begin{split}
    \phi''  = & \frac{1}{4 A_0 r (r-m)^3 \phi^2}\Bigg[8 r^3 \phi' \left(8 \pi  \mathcal{B} r^2-A_0 \phi^2+4 \pi  r^2 p\right) \\
    & -4 r^2 m \left(\phi' \left(A_0 r \phi  \phi'-4 A_0 \phi^2+16 \pi  \mathcal{B} r^2\right) +2 \pi  r p \left(4 r \phi'+\phi \right)\right) \\
    & +A_0 m^2 \phi  \left(8 r^2 \phi'^2-12 r \phi  \phi'-\phi^2\right)+4 A_0 m^3 \phi \, \phi' \left(\phi -r \phi'\right) \Bigg].
  \end{split}
\end{equation}
In all of the above expressions, the first derivative $\nu'$ has been substituted from Eq. \eqref{eqcont}. It can be seen that the set of Eqs. \eqref{masseq}-\eqref{phieq}, together with the equation of state \eqref{eqofstate}, and Eqs. \eqref{eqcont} and \eqref{fieldgamma2}, is enough to satisfy all of the field equations. 

We integrate the generalized structure equations throughout the star from the center to the surface imposing at the center appropriate initial conditions as follows:
\begin{equation}
  m(0) =0 , \quad p(0)=p_c, \quad \phi(0) = \phi_c, \quad \phi'(0)=0 ,
\end{equation}
where $p_c$ and $\phi_c$ represent the central values for the pressure and the scalar field respectively.  

\begin{figure}[h]
\centering
\includegraphics[scale=1]{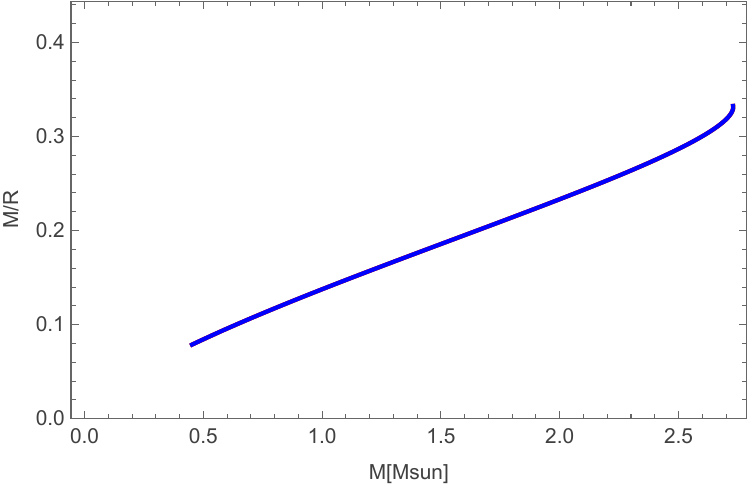}
\caption{Factor of compactness, $C=M/R$, as a function of the stellar mass.}
\label{plot1}
\end{figure}

\begin{figure}[h]
\centering
\includegraphics[scale=1]{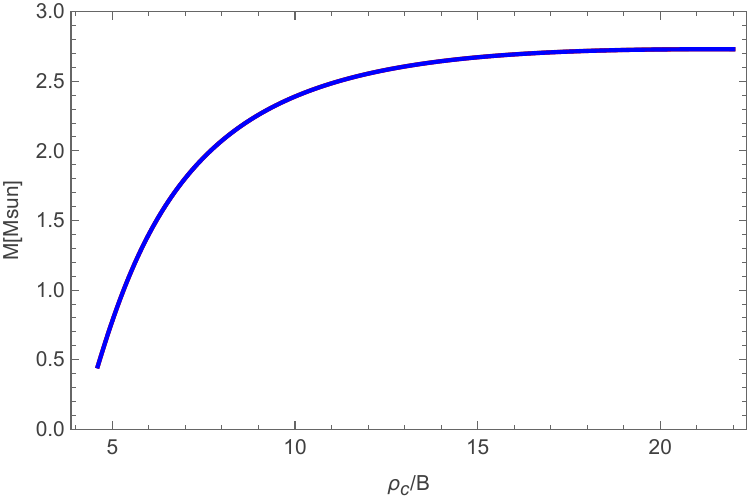}
\caption{Stellar mass as a function of the central energy density.}
\label{plot2}
\end{figure}

\begin{figure}[h]
\centering
\includegraphics[scale=1]{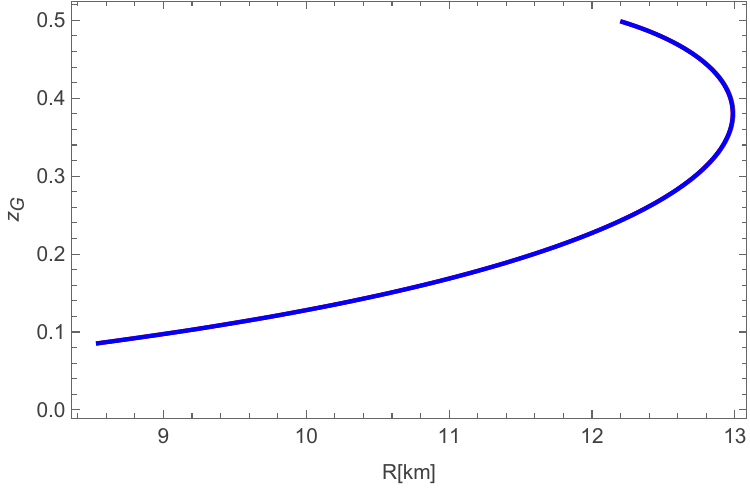}
\caption{Gravitational redshift, $z_G$, versus stellar radius $R$.}
\label{plot3}
\end{figure}

\begin{figure}[h]
\centering
\includegraphics[scale=1]{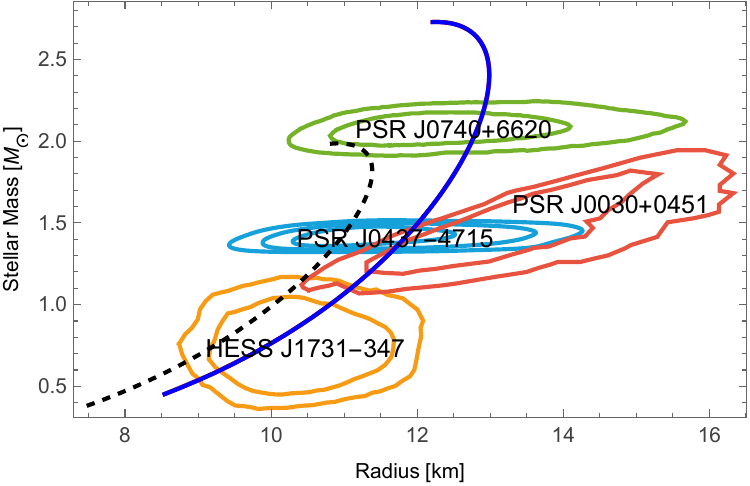}
\caption{Mass-to-radius relationships for the 3 models considered here, see text. The black dashed line corresponds to General Relativity. We have included the following astrophysical constraints: a) HESS compact object, b) NICER results, c) the most massive pulsars at two solar masses, and d) the {\bf GW}190814 event. The M-R profile passes from all allowed contours, while at the same time it accommodates massive stars at 2 solar masses and also at 2.5 solar masses.}
\label{plot4}
\end{figure}

The set of Eqs. \eqref{masseq}-\eqref{phieq} is of course quite involved and thus we approach this problem numerically. We investigate three possible configurations of parameters: a) $A_0=100$, $\phi(0)=0.1143$, b) $A_0=105$, $\phi(0)=0.1115$ and c) $A_0=95$, $\phi(0)=0.11723$, whose results are depicted in Figs. \ref{plot1}-\ref{plot4}. The values are chosen so that the value of the effective gravitational constant in the theory, $G_{eff}\sim 1/(A_0\phi^2)$ becomes asymptotically equal to Newton's constant $G_N$. We notice that the pairs of $A_0$ and $\phi(0)=\phi_c$ achieving this result in the same functional behavior (same line in the graphs) irrespectively of the particular values of $A_0$ and $\phi_0$ of the three models. We also note that the obtained mass-to-radius profile is in agreement with current astrophysical constraints, such as NICER results and the most massive pulsars.

In Fig. \ref{plot1} we present the compactness, $M/R$. As previously stated, the radius is calculated at the first root of $p(R)=0$, and the total mass is of course given by $M=m(R)$. The graph in this plot is depicted up to the point of stable configurations. As far as stability is concerned, Fig. \ref{plot2} presents the stellar mass as a function of the central energy density. Clearly, the condition
\begin{equation}
  \frac{d M}{d \rho_c} >0 ,
\end{equation}
indicating stable configurations with regards to the Harrison–Zel’dovich–Novikov criterion \cite{Harrison, ZN} is satisfied. 

In Fig. \ref{plot3} we explore the behavior of the gravitational redshift for the models under consideration. The former is calculated as
\begin{equation}\label{gravred}
  1+z_G = \frac{1}{\sqrt{-g_{tt}}} = \frac{1}{1-\frac{m(R)}{R}} .
\end{equation}
In Fig. \ref{plot4} we provide the graphs of the function of mass $M=m(R)$ with respect to the radius. For reasons of comparison, we additionally included the corresponding curve from GR (dashed line). It is apparent that the derived solutions are well inside the observational limits of known stellar configurations. 

\begin{figure}[h]
\centering
\includegraphics[scale=1]{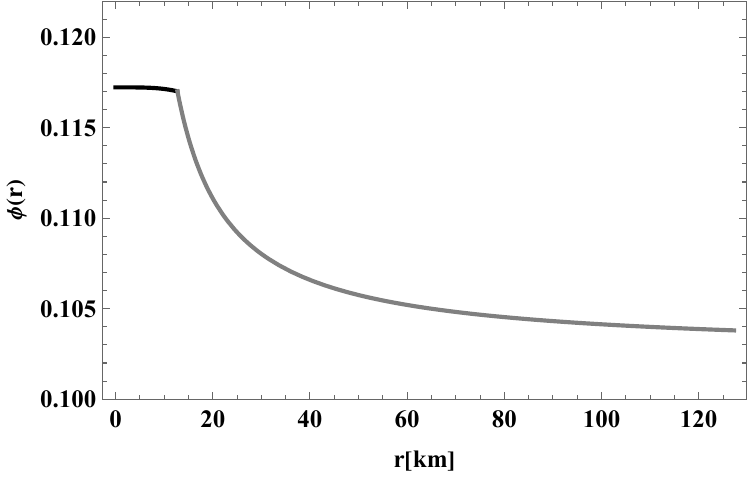}
\caption{The functional behavior of the scalar field with respect to the radial distance. The scalar field has a finite slightly decreasing value in the interior of the star and a faster decreasing value outside.}
\label{plot5}
\end{figure}

\begin{figure}[h]
\centering
\includegraphics[scale=1]{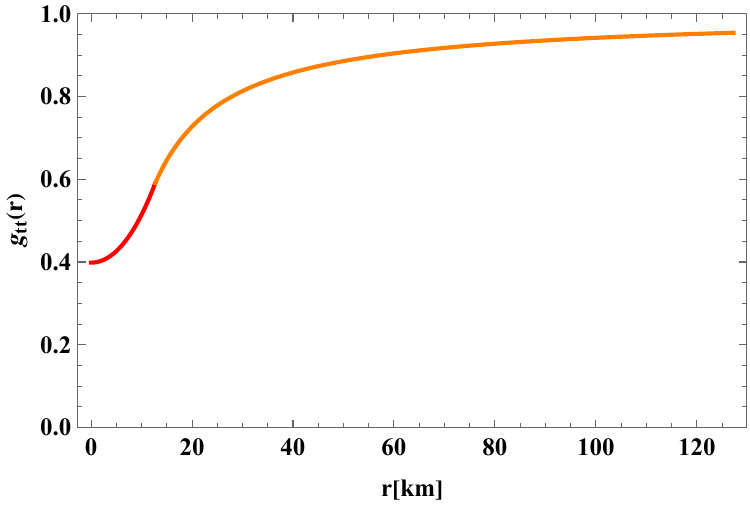}
\caption{The behavior of the absolute value of the temporal component of the metric with respect to the radius.}
\label{plot6}
\end{figure}

\begin{figure}[h]
\centering
\includegraphics[scale=1]{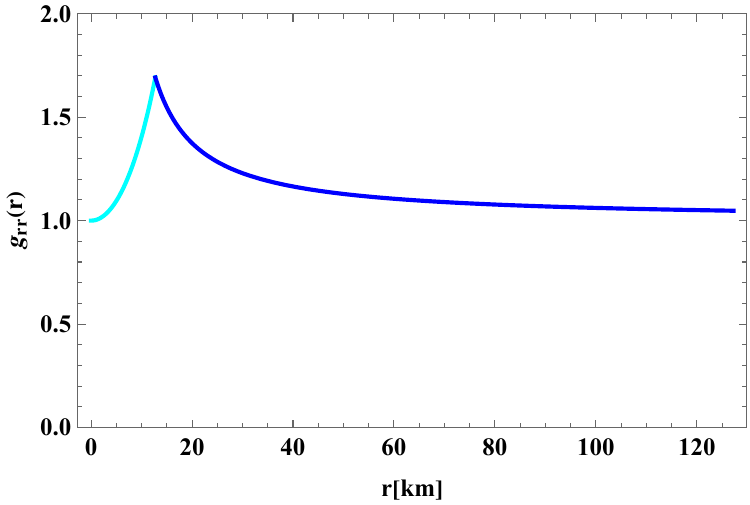}
\caption{The behavior of the radial component of the metric with respect to the radius.}
\label{plot6b}
\end{figure}

\begin{figure}[h]
\centering
\includegraphics[scale=1]{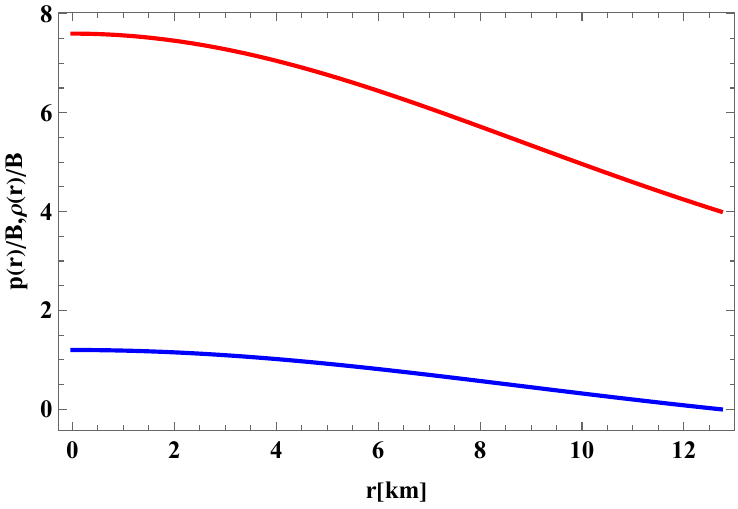}
\caption{The energy density and the pressure of the stellar fluid with respect to the radial distance.}
\label{plot7}
\end{figure}

For completeness we include the graphs of the scalar field for both the interior and the exterior solutions, Fig. \ref{plot5}, and the metric potentials Fig. \ref{plot6} and Fig. \ref{plot6b}. Finally, in Fig. \ref{plot7} we include the pressure and energy density in the interior of the star. A few comments are in order in what regards the discontinuities in the first derivatives that we observe in the plots. The equation of state under consideration Eq. \eqref{eqofstate} leads to a discontinuous energy density at the border of the stellar radius, $r=R$. To see this, take the pressure which is a continuous function that vanishing at the border, $p(R)=0$. Due to Eq. \eqref{eqofstate}, the energy density is $\rho(R)=4\mathcal{B}\neq0$. However, in the vacuum, for $r>R$, we have a finite jump to $\rho=0$. This, unavoidably leads to a discontinuous first derivative for the $g_{rr}$ component (the $\lambda'$ in terms of the potential) even in the case of general relativity. In this latter case, the radial equation would also dictate continuity for the other metric potential derivative, the $\nu'$. However, in the theory under consideration, by taking the limits of the radial equation from the left and from the right of the stellar radius we obtain the relation (for a continuous zero pressure on shell):
\begin{equation} \label{nuphi}
  \nu'(R_+) - \nu' (R_-) = \frac{4R}{\phi(R)^2} \left( \phi'(R_+)^2 -\phi'(R_-)^2 \right).
\end{equation}
In principle, the $\nu'$ and $\phi'$ can both be discontinuous. Moreover, if the discontinuities satisfy the above relation, a thin shell with a radial pressure is not necessary to account for the observed discontinuity. Note that this cancelation is fundamental as the radial equation corresponds to the Hamiltonian constraint and should balance out any discontinuities in the system of equations.

\begin{figure}[h]
\centering
\includegraphics[scale=1]{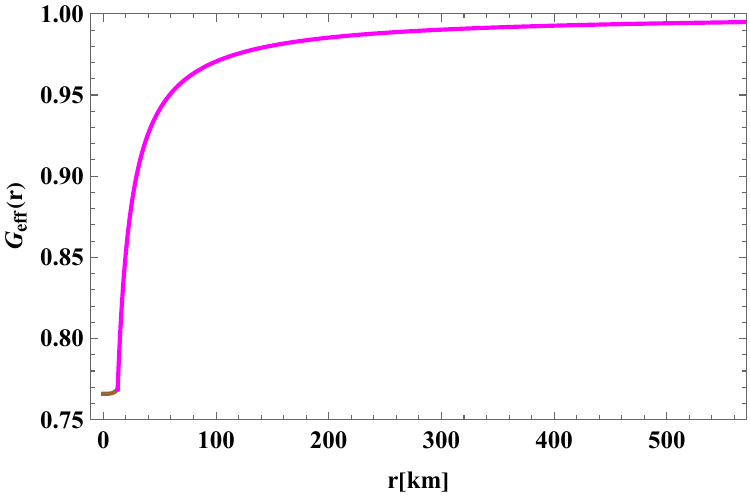}
\caption{The effective gravitational constant as a function of the radial distance. The $G_{eff}$ is measured in units of $G_N=1$. We see that it asymptotically approaches the values of the Newtonian constant.}
\label{plot8}
\end{figure}

From Fig. \ref{plot5}, we observe that the scalar field slightly decreases in the interior solution, and continues to do so in the outside. The latter is also apparent from  \eqref{a0}. Due to the form of the theory we consider here, we note that the effective Newtonian constant falls with the square of the scalar field, $G_{eff}\sim 1/\phi^2$, thus implying that $G_{eff}$ is greater outside of the star and smaller in the inside. In Fig. \ref{plot8}, we depict the exact behavior of the effective gravitational constant $G_{eff}$ as a function of the radius, as measured in geometric units ($G_N=1$).

As a final remark we should add that nonmetricity theories, like $f(Q)$ gravity are known to harbor ghost degrees of freedom \cite{ghost1,ghost2}. Although it may happen that such degrees of freedom might not propagate \cite{ghost3} or, under certain conditions, not necessarily cause instabilities \cite{ghost4}, this raises reasonable questions on how healthy a theory involving nonmetricity may be. The scalar-tensor nonmetricity theory is not equivalent to $f(Q)$, unless $B(\phi)=0$ in the action \eqref{action}. So, having $B\neq0$ makes a direct comparizon with the results of $f(Q)$ highly non-trivial. Nevertheless, at least for the particular configuration with which we deal here, we have some strong indications of a healthy theory. To see this let us turn to the original field equation Eq. \eqref{fieldphi} for the scalar field. The theory we consider is characterized by $B=-8A_0$, $A=A_0 \phi^2$ and $V(\phi)=0$, which leads to a scalar equation of the form $\phi''-\phi Q/8=0$. The definition of the nonmetricity scalar we use, \eqref{Qscdef}, leads to a $Q$ being everywhere negative, while $\phi$, as can be seen from Fig. \ref{plot5} and \eqref{a0}, is positive both for the interior and the exterior solution. As a result we have an equation of the form $\phi''+ dV_{eff}/d\phi=0$, where $dV_{eff}/d\phi$ is positive. Thus, we can say that the scalar field goes down the potential towards a minimum. Finally, the finiteness of the value of the scalar field everywhere is also a strong indication of a well-behaved theory.

%%%%%%%%%%%%%%%%%%%%%%%%%%%%%PLOTS%%%%%%%%%%%%%%%%%%%%%%%%%%%%

%%%%%

%%%%

%%%%

%%%%

%%%%%%%%%%%%%%%%%%%%%
\section{Discussion and Concluding remarks}
%%%%%%%%%%%%%%%%%%%%%

To summarize our work, in the present article we have investigated the existence of viable compact objects within the framework of symmetric teleparallel scalar--tensor gravity. In particular, we have analyzed the physical properties of interior solutions with an exterior extremal Reissner--Nordstr\"{o}m geometry. We have adopted a symmetric and flat connection, which
allowed us to explore nontrivial gravitational solutions beyond the limit of STEGR. 

The field equations of the proposed gravitational model are of second-order and admit a minisuperspace description. We thus applied the method of variational symmetries to a reduced mechanical system and derive the corresponding conservation laws for the gravitational equations. We used the symmetry criterion as a selection rule to distinguish the theory with a quadratic nonminimal coupling to the nonmetricity scalar. Within this formulation, we successfully recovered previous results on vacuum solutions and demonstrate that their existence originates from variational symmetries of the minisuperspace Lagrangian. Interestingly enough, although the symmetric teleparallel scalar-tensor framework includes $f\left( Q\right) $-gravity as a special case, the specific solutions do not exist in the $%
f\left( Q\right) $ limit. This is the consequence of the solutions being derived for a vanishing potential function for the scalar field. 

Concerning the interior solutions, we adopted a particular equation-of-state and calculated numerically several quantities of interest including: stellar mass and radius, the factor of compactness and the gravitational redshift. In addition to that, the Harrison–Zel'dovich–Novikov criterion suggests that the constructed configurations are stable, and we observed that it nicely fits current astrophysical constraints. What is more, the factor of compactness remains smaller than $4/9$ in agreement with the Buchdahl limit of GR \cite{Buchdahl}. From this analysis, we conclude that the symmetric teleparallel scalar--tensor model can successfully describe local gravitational phenomena. In a future work, we plan to implement additional astrophysical tests in order to further elaborate on the viability of the theory.

%%%%%%%%%%%%%%%%%%%%%%%%%%
\begin{acknowledgments}
A.~P  thanks the support of VRIDT through Resoluci\'{o}n VRIDT No. 096/2022 and Resoluci\'{o}n
VRIDT No. 098/2022. Part of this study was supported by FONDECYT 1240514.
\end{acknowledgments}
%%%%%%%%%%%%%%%%%%%%%%%%

\end{document}